# Breather Soliton Dynamics in Microresonators

Mengjie Yu[1,2], Jae K. Jang[1], Yoshitomo Okawachi[1], Austin G. Griffith[3]

Kevin Luke[3], Steven A. Miller[2,4], Xingchen Ji[2,4], Michal Lipson[4], and Alexander L. Gaeta[1] [*]

[1]*Department of Applied Physics and Applied Mathematics, Columbia University, New York, NY 10027*

[2]*School of Electrical and Computer Engineering, Cornell University, Ithaca, NY 14853*

[3]*School of Applied and Engineering Physics, Cornell University, Ithaca, NY 14853*

[4]*Department of Electrical Engineering, Columbia University, New York, NY 10027*

**The generation of temporal cavity solitons in microresonators results in low-noise optical frequency combs which are critical for applications in spectroscopy, astronomy, navigation or telecommunications. Breather solitons also form an important part of many different classes of nonlinear wave systems with a localized temporal structure that exhibits oscillatory behavior. To date, the dynamics of breather solitons in microresonators remains largely unexplored, and its experimental characterization is challenging. Here, we demonstrate the excitation of breather solitons in two different microresonator platforms based on silicon nitride and on silicon. We investigate the dependence of the breathing frequency on pump detuning and observe the transition from period-1 to period-2 oscillation in good agreement with the numerical simulations. Our study presents experimental confirmation of the stability diagram of dissipative cavity solitons predicted by the Lugiato-Lefever equation and is importance to understanding the fundamental dynamical properties of solitons within the larger context of nonlinear science.**

Temporal cavity solitons (CS's) are self-localized, pulses of light that can be excited in nonlinear optical resonators [1-8] and have recently attracted significant research interest in the context of microresonator-based frequency comb generation [5-18]. In contrast, the breather soliton presents a nonlinear wave in which energy is localized in space but oscillates in time, or vice versa, and is found in a wide variety of subfields of physics such as solid-state physics, fluid dynamics, plasma physics, molecular biology, and nonlinear optics. Here, we present a theoretical and experimental study of breather CS's [19-22] excited in microresonators. This work provides the first experimental observation and characterization of such dynamic instabilities in optical microresonators [18,20-24], the understanding of which is critical to the microresonator-based frequency combs and is relevant to a large variety of physical systems for both fundamental and applied interests. We demonstrate the universal nature of such breather solitons in two different material platforms, silicon nitride ($Si_3N_4$) [9,10,12,13] and silicon (Si) [16,17] and for two distinct spectral regimes. Our results establish a direct link between the breathing frequency and the pump-cavity detuning which is used to excite temporal CS's in microresonators and also show that the breather solitons can oscillate at frequencies well outside of the cavity resonance linewidth.

Temporal CS's were first studied experimentally in fiber cavities [3,20,25,26] and subsequently reported in optical microresonators [4-6]. Mathematically, temporal CS's are a steady-state localized solution of the Lugiato-Lefever equation (LLE; see Methods) [27] which has been extensively used to model driven nonlinear optical resonators [2,19-24,28-30]. The LLE allows for modeling of rich cavity dynamics, including spatiotemporal instabilities [20,21,31] and in particular, predicts the existence of persistent breathing CS's in optical resonators pumped by a continuous-wave (cw) laser, as experimentally verified in a fiber cavity

[20].

In Figs. 1(a) and 1(b), we simulate temporal evolution of a breathing temporal CS and the evolution of the CS peak power, respectively, in a microresonator using the LLE. It can be seen that as the CS evolves, it undergoes a periodic oscillation in time, which is accompanied by the associated breathing of the CS spectral envelope with the same period [22]. The optical spectrum and the comb transmission as the pump laser frequency is swept through a cavity resonance [4,26] are shown in Figs. 1(c) and (d) (see Methods). When the pump laser is tuned across the resonance starting from the effectively blue-detuned regime [4,26], the intracavity field initially develops into a low-noise primary comb state [e.g. Fig. 1(c)(i)], destabilizes to a high RF amplitude noise state [Fig. 1(c)(ii)], and eventually settles into the soliton state [Fig. 1(c)(iv)]. This transition into the soliton state coincides with the pump frequency crossing the effective zero detuning and is accompanied by an abrupt change in the cavity transmission [Fig. 1(d)] [4]. More importantly, we identify the breather soliton state [Fig. 1(c)(iii)] by observing a periodic oscillation in time similar to Figs. 1(a) and (b). From this analysis, we conclude that the breathing regime of CS's is located in the region of detunings between the unstable high-noise state and the stable CS regime, which is consistent with previous theoretical studies [20-24].

In our experiment, we observe breather solitons in $Si_3N_4$ microresonators in the telecommunication regime and in Si microresonators in the mid-infrared (mid-IR). In $Si_3N_4$ microresonators, the spectral evolution towards soliton formation in the optical and RF domain is plotted in Fig. 2 (a)(i)-(iv). These plots, from top to bottom, correspond to the generation of primary comb lines, multiple mini-combs, breather solitons, and stable solitons. Soliton formation is confirmed by low RF noise [Fig. 2 (a)(iv)] and the abrupt cavity transmission step shown in Fig. 2(b). The time evolution of the comb output for the breather soliton state is recorded in Fig. 2(c) showing the temporal oscillatory behavior with a time period of 7 ns, which corresponds to an RF beat note of 155 MHz. The excitation of breather solitons in both platforms requires a certain range of pump power and can be achieved by tuning the pump frequency into resonance without any additional pump power modulation [22].

In our Si microresonators, we monitor and analyze different states of the generated frequency combs by measuring the three-photon-absorption (3PA) induced photocurrent extracted from the integrated PIN junction, which is an effective way to probe the temporal behavior inside a cavity [17] (see Methods). Figures 2(d)(i)-(iv) shows the evolution of the generated optical and RF spectra as the pump laser is tuned into resonance. Initially, from Fig. 2(d)(i) to (ii), the DC component of the 3PA-induced current gradually increases due to intracavity power build-up. These spectra correspond to high-noise states in which multiple mini-combs grow and interact with each other leading to strong and broad RF beat notes [12]. Next, we observe a transition to a state with a more structured optical spectrum and an abrupt increase in the measured DC current from 1.199 mA to 1.683 mA [Fig. 2(d)(iii)]. The corresponding RF spectrum shows a sharp peak at 336 MHz with a low noise background. Tuning the pump frequency further leads to a low-noise RF spectrum without any peaks [Fig. 2(d)(iv)]. The variation in the DC current with pump detuning is shown in Fig. 2(e). The sudden increase in the DC current indicates soliton formation due to its higher peak power (see Methods). We further confirm this by observing the concurrence between the abrupt current increase and a cavity transmission step [3,5]. Thus, the state with the observed sharp RF peak [Fig. 2(d)(iii)] corresponds to a breather soliton state that exists beyond the high-noise state but before the stable soliton state [Fig. 2(d)(iv)], which agrees with our simulations and past theoretical studies [4,20,21]. Here, the soliton breathes at a rate of 336 MHz, which is within the cavity linewidth and corresponds to a period of 372 cavity roundtrips. The narrow-linewidth RF beat note corresponding to the breather soliton state can be distinguished from the sharp RF beat notes observed at the mini-comb generation stage [13] since each one occurs at different stages of comb evolution. The two states have drastically different temporal behavior (*e.g.* pulse formation in breather soliton state but not in mini-comb state), which is easily detected by our technique here by measuring the 3PA-induced photocurrent.

Another important signature of breather soliton dynamics is that when transitioning into the breather soliton state, the RF beat note corresponding to the breather frequency becomes significantly narrower while maintaining the same center frequency, as shown in Fig. 2 (a). Such evolution is also observed in some Si microresonators for which the higher order harmonics of the 3PA-induced current noise do not obscure this feature. This suddenly narrowed RF beat note is also well predicted by our numerical simulations of the RF spectra of the comb transmission at different comb states corresponding to Fig. 1(c). Our modeling indicates that the breathing CS manifests itself through low-frequency (relative to the comb spacing) modulation sidebands whose frequency offsets from each comb line correspond to the breathing period [18], shown in Fig. 3 (green curve). In addition, the simulated RF spectra at three different pump detunings [Fig. 1(c) (ii) to (iv)] are plotted in Fig. 3. It can be seen that at the detuning $\Delta = 3.9$, [(ii); blue curve] the system is initially in a high-noise state characterized by a broad primary noise peak (centered at $f = 0.73$). In the time and spectral domains, the intracavity field fluctuates significantly from one roundtrip to another, analogous to spatiotemporal chaos originally discussed in the context of plasma physics [31]. In contrast, as the detuning is increased past a threshold value, the RF noise peak abruptly sharpens while maintaining its central frequency [(iii) $\Delta = 4.5$; green curve]. We plot the coupled Lorentzian resonance curve of the resonator under consideration on the same axes (right axis; red curve). Note that the simulated breathing frequency is 2.3 times the resonance linewidth. The sideband eventually drops to a negligible level with an increased detuning [(iv) $\Delta = 8$; black curve] which signifies the stabilization of the temporal CS's, where the pulses reach a time-stationary state.

We further characterize the breather dynamics and investigate on what factors the breathing frequency is dependent. We observe the breathing frequency can be excited at a value of multiple cavity linewidth; we observe a beat note of 3.7 GHz in a Si microresonator and of 390 MHz in a $Si_3N_4$ microresonator, both of which are in the range from 3 to 4 times its cavity linewidth. Remarkably, as our numerical simulation in Fig. 3 revealed, the modulation sideband associated with the breathing dynamics can also oscillate well outside the resonance linewidth. This occurs since the soliton breathing manifests itself as a collective amplitude modulation of all the comb lines at the same frequency [18], resulting in a larger gain at that frequency even if it is outside of the resonance linewidth. In addition, we find that the breathing frequency can be continuously tuned by changing the pump laser detuning. By tuning the pump to shorter wavelength while the soliton state still persists, we observe an increase in oscillation frequency as well as a decrease in modulation depth in both platforms as shown in Fig. 4. In the Si microresonator, it is also accompanied by a decrease in DC current corresponding to a decreasing soliton peak power resulted from a smaller pump-cavity detuning. This trend agrees with the numerical studies in [18,21]. Our simulation also shows this dependence of the breathing frequency on the pump-cavity detuning, and we find that as the pump detuning is varied, both the intensity and frequency of the RF beat note are observed to change continuously, whereas those in the mini-comb stage can change abruptly [13].

Microresonators allow for assessing larger values of the detuning ($\Delta$) because of its high finesse as compared to its fiber-cavity counterpart. At higher values of the detuning, more complicated dynamical regimes can be found in the LLE for higher values of the pump power, such as period-2 oscillations, period-3 oscillations, period-N oscillations, temporal chaos and spatial-temporal chaos [20]. In the $Si_3N_4$ microresonator, using a higher pump power than that in Fig.2 (a), we observe a transition from the period-1 oscillation to the period-2 oscillation as shown in Figs. 5 (i-ii), and it finally stabilizes to the stable CS's [Fig. 5(iii)] as increasing the pump detuning. It is the first time that an oscillatory temporal CS's with higher periodicity has been observed experimentally and agrees with the previous theoretical prediction [20].

In conclusion, we show the universality of the dynamics of the breather CS formation in microresonators, attributing the formation of narrow low-frequency RF modulation sidebands in the modelocked regime to time-oscillating behavior of CS's. The universal nature of breather soliton formation is indicated by our its observation in two distinct platforms that have different material characteristics, device geometries and operating conditions. Our results present experimental confirmation of the pioneering theoretical studies of a breather soliton solution in a cw-pumped high-finesse microresonator and provide experimental techniques for their observation, which agrees well with our numerical results. Our work also establishes a direct link between the breathing frequency and the pump-cavity conditions, and reveals microresonators can be an ideal test bed for fundamental theories of nonlinear wave dynamics that are relevant to a large variety of physical systems.

**Methods**

**Numerical simulations**

A variety of spatiotemporal dynamics in driven passive resonators can be modeled based on a modified Lugiato-Lefever equation (LLE)

$$t_R \frac{\partial E(t,\tau)}{\partial t} = \left[ -\alpha - i\delta_0 + i\gamma L |E|^2 + iL \sum_{k \geq 2} \frac{\beta_k}{k!} \left( i\frac{\partial}{\partial \tau} \right)^k \right] E(t,\tau) + \sqrt{\theta} E_{in}.$$

Here $E(t,\tau)$ is the slowly varying intracavity field envelope, $t$ is the slow evolution time on the order of the cavity roundtrip time $t_R$, while $\tau$ is a fast time traveling at the group velocity of the light inside the resonator. $\alpha$ is a fraction corresponding to half the total percentage power loss per roundtrip, and $L$ is the cavity length. $\gamma$ is the nonlinear parameter, and $\beta_k$ is the kth order dispersion coefficient. Finally, $\theta$ is the input coupling coefficient, $E_{in}$ is the pump field, and $\delta_0$ is the phase detuning of the pump with respect to the nearest cavity resonance. To ensure broader applicability of our study, we employ the following dimensionless form of the LLE

$$\frac{\partial F(t',\tau')}{\partial t'} = \left[ -1 - i\Delta + i|F|^2 + i\sum_{k \geq 2} d_k \left( i\frac{\partial}{\partial \tau'} \right)^k \right] F(t',\tau') + S,$$

where we have used the same normalization convention as that in [3,24], with the additional expression for the normalized dispersion coefficients $d_k = \frac{\beta_k L}{\alpha(k!)} \left( \frac{2\alpha}{|\beta_2| L} \right)^{k/2}$, such that the 2nd order dispersion coefficient $d_2$ represents the sign of the group-velocity dispersion ($d_2 = -1$ in our case). The optical frequency $f$ and RF frequency $\nu$ used in Fig. 1 correspond to the fast time $\tau'$ and slow time $t'$, respectively. Note that $S$ and $\Delta$ are the normalized pump field strength and detuning, respectively, and are the only control parameters of the current system under consideration, thus greatly simplifying our investigation. For specificity, however, we use parameters similar to our 200 GHz free-spectral-range (FSR) Si$_3$N$_4$ resonator with 950×1500 nm cross-section, i.e. $t_R$ = 1/FSR = 5 ps, $\alpha$ = 0.0018, $\theta$ = 0.0005, $\gamma$ = 0.9 W$^{-1}$m$^{-1}$, $L$ = 0.63 mm, $\beta_2$ =

-202 ps²/km, $\beta_3$ = 0.034 ps³/km, $\beta_4$ = 7.7 x 10⁻⁴ ps⁴/km, and $|E_{in}|^2$ = 170 mW. Depending on the simulation, we vary $\delta$ from -0.005 rad to 0.022 rad. These parameters correspond to $|S|^2$ = 8.7 and $\Delta$ = -2.9 to 12, $d_2$ = -1, $d_3$ = 0.003, and $d_4$ = 9 x 10⁻⁶. Although we include higher-order (3rd and 4th) dispersion coefficients for completeness, their effects are negligible and do not invalidate the applicability of the past theoretical studies to our case.

The LLE is integrated using the split-step Fourier method. In Fig. 1(c) and (d), we seed the initial parametric process with the noise consisting of a photon per mode with random phase. The normalized pump power $|S|^2$ = 8.7 is used and the detuning $\Delta$ is varied from -2.9 and 12 at a rate of 0.001 per one normalized slow time unit. The RF spectra in Fig. 3 are obtained by using the intracavity fields saved during the detuning scan simulation as the initial conditions, and regularly sampling the roundtrip-averaged out-coupled power $|E_{out}|^2$ (measured in W), where $E_{out} = \sqrt{\frac{\alpha\theta}{\gamma L}}\left(E - \frac{\alpha}{\theta}S\right)$. For each detuning, before sampling, we numerically propagate the intracavity field for a sufficiently long time (150 slow time units, or approximately 300 cavity photon lifetimes) to allow the system to reach steady-state (quasi steady-state for breather soliton and high-noise states). For Figs. 1 (a) and (b), we use the fixed pump power and detuning ($|S|^2$ = 8.7, $\Delta$ = 4.5), and an approximate CS solution $E(\tau) = \sqrt{2\Delta}\,\text{sech}(\sqrt{\Delta}\tau)$ as the initial condition.

**Devices and experimental set-ups.** We utilize a high-$Q$ etchless Si microresonator with a device structure as described in [16] with an FSR of 127 GHz. The resonator is pumped using a cw optical parametric oscillator (<100 kHz linewidth). Taking advantage of the inherent free-carrier (FC) effect in silicon, an experimental setup is used as described in [17]. The measurement is taken by tuning the pump laser centered at 3.07 μm into a cavity resonance with an off-resonance bus waveguide power of 80 mW. In order to reduce the optical losses due to FC generated from three-photon absorption (3PA), we use an integrated PIN junction and apply a reverse bias voltage of -12 V [13]. Simultaneously, the DC and RF components of the current extracted from the PIN are monitored using a Keithley SourceMeter and a RF spectrum analyzer, respectively. The measured RF current reflects noise properties of intracavity comb generation with a bandwidth of 12 GHz, which in this case is limited by the RF amplifier we use before the RF spectrum analyzer. More importantly, due to the intrinsic 3PA process, the FC is generated at a rate proportional to the cube of the optical intensity instantaneously, yet with a decay rate determined by the FC lifetime of a specific PIN structure, expressed in the equation below taking into account the boundary condition of the microresonators:

$$\frac{\partial N_c(t,\tau)}{\partial \tau} = \frac{\beta_{3PA}}{3\hbar\omega} \frac{I(t,\tau)^3}{A_{eff}^3} - \frac{N_c(t,\tau)}{\tau_{eff}}$$

where $N_c(t,\tau), \beta_{3PA}, I(t,\tau), \tau_{eff}, A_{eff}$ are the FC density, 3PA coefficient, optical intensity, FC lifetime and the effective mode area, respectively while $\omega$, $t$, and $\tau$ are the pump frequency, and fast and slow time scales, respectively. Typically, in a high-noise comb state, instantaneous intracavity optical power is relatively low and features quasi-cw behavior due to a low spectral coherence while a soliton state exhibits short pulses with high peak power on top of a low cw background within one roundtrip. Therefore, an increased 3PA-induced current is always expected at the transition to soliton states even though the average intracavity power decreases in an ultra-compact microresonator which enables broadband comb operation. Additionally, any fluctuation of intracavity power will result in the fluctuation of current with the fundamental frequency, along with its second and third order harmonics. The fundamental frequency modulation depth will be enhanced by the square of the DC component of the current. This novel technique yields much higher sensitivity and larger bandwidth to measure the temporal oscillation of CS's. In this sense, the 3PA-induced current is a powerful tool to probe the temporal behavior of the generated frequency combs, especially in the soliton modelocked regime which features pulse formation with much higher peak power. The detection of the breather soliton state requires fast and sensitive measurement techniques, especially when the oscillating frequency is far off the resonance or the modulation depth is not significant. Our technique based on the 3PA-induced current is experimentally proven to be an efficient way to observe these dynamics.

For comb generation in the $Si_3N_4$ microresonator, which has a cross-section of 950×1500 nm and an FSR of 200 GHz, we detune the pump laser at 1540 nm into a cavity resonance with a bus waveguide power of 56 mW. The $Q$ factor is estimated to be more than 2 million. To measure the RF spectrum, we filter out the pump wavelength using a wavelength-division multiplexing coupler and send the output to a photodiode. Similar breather soliton dynamics is also demonstrated in $Si_3N_4$-microresonators which have a cross section of 950×1400 nm and an FSR of 500 GHz, and is pumped at 1560 nm. The measured $Q$ factor is 1.9 million.

**Acknowledgment**

We acknowledge support from Defense Advanced Research Projects Agency (W31P4Q-15-1-0015), the Air-Force Office of Scientific Research (FA9550-15-1-0303), and National Science Foundation (ECS-0335765, ECCS-1306035). This work was performed in part at the Cornell Nano-Scale Facility, a member of the National Nanotechnology Infrastructure Network, which is supported by the National Science Foundation (NSF) (grant ECS-0335765).

**Author contributions**

M.Y. and J.K.J. prepared the manuscript in discussion with all authors. M.Y., J.K.J., and Y.O. designed and performed the experiments. K.L., S.A.M. and X.J. fabricated the devices. J.K.J. performed theoretical modeling and M.Y. performed preliminary numerical simulations. M.L. and A.L.G. supervised the project.

M.Y. and J.K.J. contributed equally to this work.

**Additional information**

The authors declare no competing financial interest.

**Corresponding author**

Correspondence: Author to whom all correspondence should be directed.

Alexander L. Gaeta – alg2207@columbia.edu

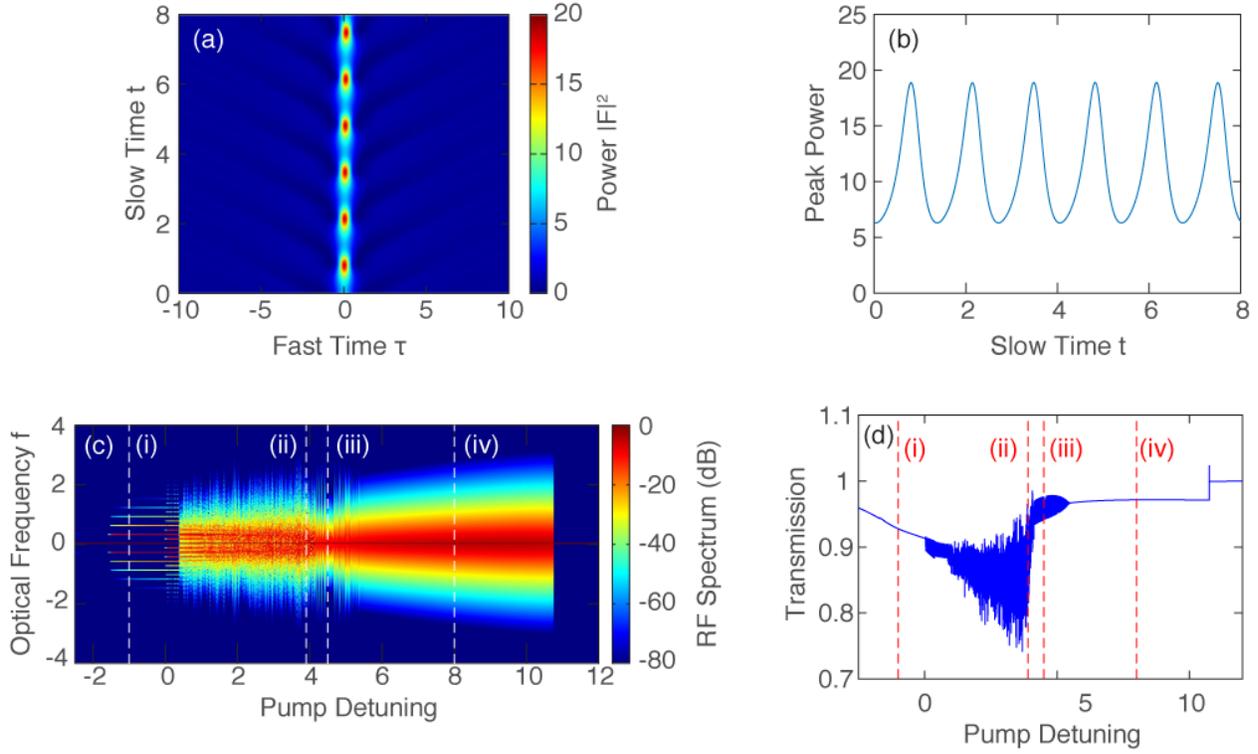

**Figure 1 | Numerical simulation of breather solitons.** (a) The temporal evolution of a breathing temporal cavity soliton(CS). The period of the breathing cycle is 1.34 time units. (b) Plot of the temporal evolution of the peak power of the CS. (c) Simulated density plot of instantaneous optical spectra as a function of the pump detuning and (d) the corresponding normalized resonator transmission. The vertical dashed lines, from left to right, correspond to representative examples of (i) the primary comb state ($\Delta = -1$), (ii) high-noise state ($\Delta = 3.9$), (iii) breather soliton state ($\Delta = 4.5$), and (iv) stable soliton state ($\Delta = 8$).

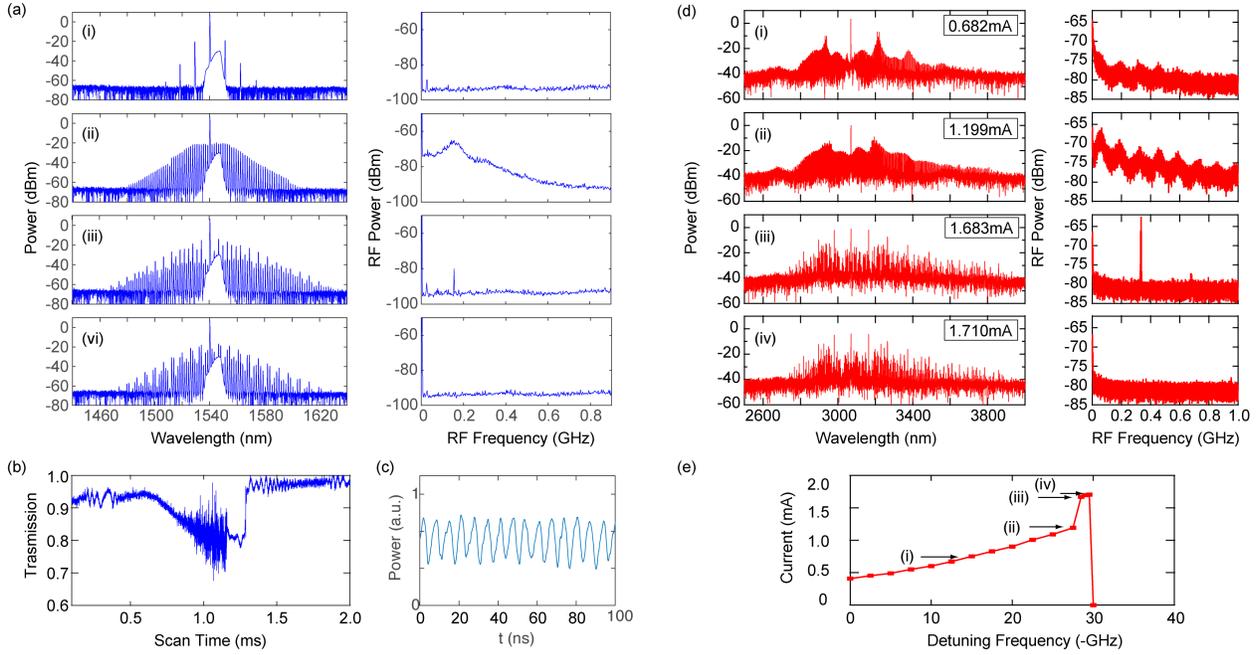

**Figure 2 | Observation of breather solitons in $Si_3N_4$ (a-c) and Si (d, e) microresonators** (a) Optical (left) and RF (right) spectral evolution showing breather solitons in $Si_3N_4$. (b) The pump power transmission as the laser frequency is scanned across the resonance. The transmission step is indicative of soliton formation. (c) The recorded time trace of the comb output power at the breather soliton state (iii). In a Si microresonator: (d) Optical (left) and RF (right) spectral evolution showing breather solitons in Si. Optical spectrum in (iv) spans 0.8 of an octave (from 2.4 to 4.3 μm). No other significant features are observed in the RF spectra beyond 1 GHz up to 12 GHz. (e) Three-photon-absorption (3PA) induced current measured by scanning the pump detuning. Arrows correspond to four states (i to iv) in (a), which are (i, ii) high-noise states, (iii) breather soliton and (iv) stable soliton states.
16

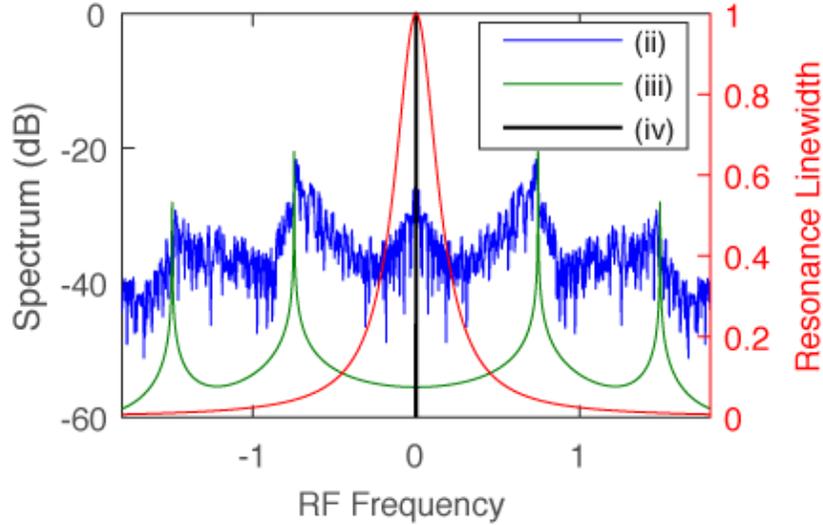

**Figure 3 | Numerical simulations of RF spectra** at selected detunings corresponding to Fig. 1 (c) (ii)-(iv). For comparison, the plot also shows the coupled Lorentzian resonance curve (red curve; right axis). Note that all the displayed quantities are dimensionless. For details on simulations, see Methods.

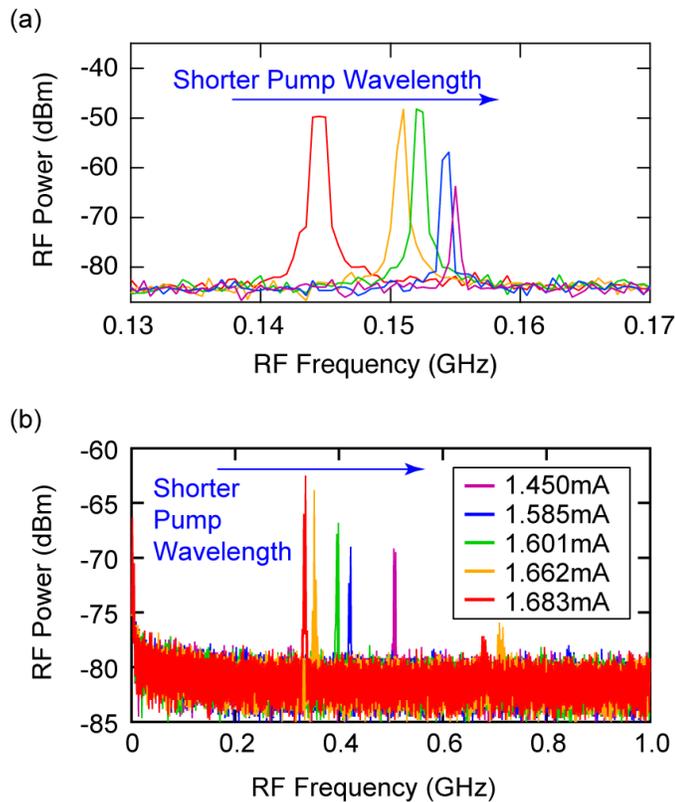

**Figure 4 | The breathing frequency at different pump-cavity detunings.** Breathing frequency measured as the pump laser is tuned to shorter wavelength closer to the resonance which corresponds to the arrow (a) in a $Si_3N_4$ microresonator and (b) in a Si microresonator at different DC 3PA-induced currents.

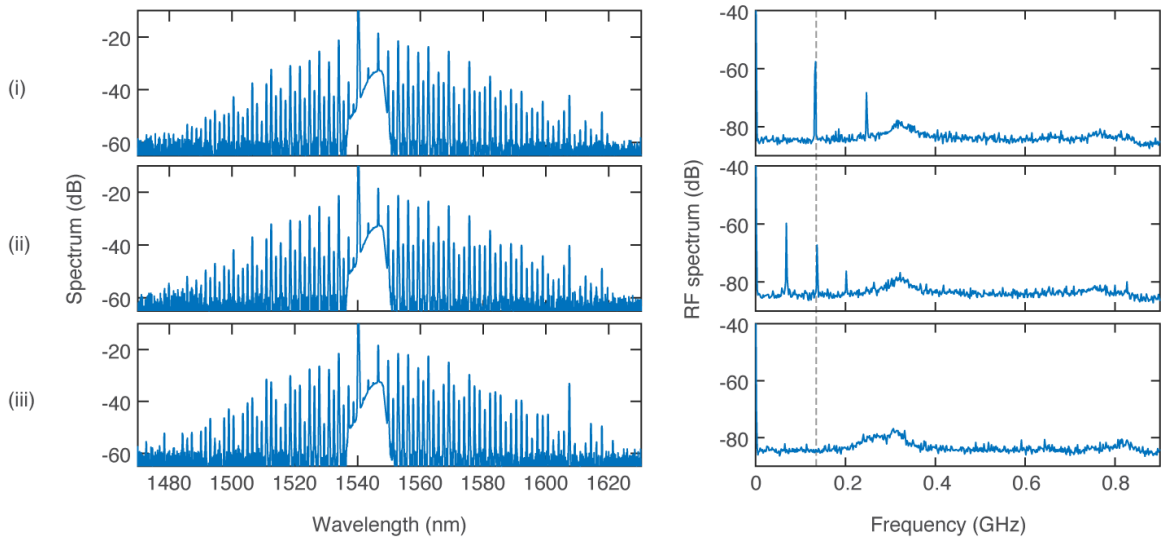

**Figure 5 | Observation of a temporal cavity soliton (CS) with higher oscillatory periodicity in a $Si_3N_4$ microresonator.** At higher power powers, as the pump detuning increases, the period-1 oscillation (i) transitions to a period-2 oscillation (ii), and the CS finally stabilizes (iii). The dash line indicates that the first RF beat note in state (i) aligns with the second RF beat note in state (ii) at the transition point.

18